\documentstyle [11pt] {article}
\def\mic{$\mu$}

\title{\bf A Study of External Galaxies Detected by the COBE Diffuse
Infrared Background Experiment}
\author{ Sten Odenwald$^1$, Jeffrey Newmark$^2$,\\ 
George Smoot$^3$ }

\begin{document}
\maketitle

\centerline{ $^1$ Hughes STX, Code 685.9, Goddard SFC, Greenbelt, MD 20771}
\centerline{ $^2$ Applied Research Corporation, Landover, MD 20785}
\centerline{ $^3$ Lawrence Berkeley Laboratory, University of California, 
Berkeley, CA 94720 } 
\newpage
\centerline{\bf ABSTRACT}

A comparison of the
 COBE\footnote{
The National Aeronautics and Space Administration/Goddard Space Flight
Center (NASA/GSFC) is responsible for the design, development, and 
operation of the {\it COBE}.
Scientific guidance is provided by the {\it COBE} Science Working Group.
GSFC is also responsible for the development of the analysis software and
for the production of the mission data sets.} 
 Diffuse Infrared Background Experiment (DIRBE) all-sky survey
 with the locations of known galaxies in the IRAS Catalog
of Extragalactic Objects and the Center for Astrophysics Catalog of
Galaxies led to the detection of as many as 56 galaxies. In this
paper, we present the photometric analysis of these detections
and an analysis of their dust properties.

The sample is dominated by late-type spirals and active galaxies, and
their far-IR continuua between 25-240 \mic m suggest a single dust component
dominating the far-IR emission. We show that
for the spirals, $< T_d > = 27.6$ K, and for the active galaxies,
$< T_d > = 35.5$ K. Estimates of the ratio of the mass of the 
dust component detected at $T_d = 20 - 30$ K to a hypothetical component
with $T_d = 15$ K provide modest constraints to the Very Cold Dust (VCD) 
component present
in the galaxies detected at 140 and 240 \mic m. VCD components with $< 7$
times the mass of the 20-30 K dust are consistent with all of the upper limits.

\vskip 0.5truein
{\it Subject Headings: Infrared:Sources, galaxies:photometry}
\newpage
\vskip 0.2truein
\centerline{\bf 1.0  INTRODUCTION }

The Infrared Astronomical Satellite (IRAS) has shown that many galaxies are
strong sources of far-IR radiation, particularly at 60 and 100 \mic m  
(Soifer, Houck and Neugebauer, 1987).  The nature of galactic far-IR emission 
among late-type spirals is widely acknowledged to be 
radiation reprocessed by dust grains present in a variety of sizes
and temperatures in the interstellar medium (Cox and Mezger, 1989; Desert et
al., 1990). This radiation may  have as its origin the UV-rich light from new 
generations of OB stars in spiral arms; the non-thermal continuum produced by 
an active nuclear source (Knapp and Patten, 1991); or the diffuse
interstellar radiation field itself (Mathis, Mezger and Panagia, 1983). 
 An open question regarding the infrared
properties of `normal' galaxies is the number of dust components
contributing to their far-IR luminosities. A perusal of the literature reveals
a confusing field of possibilities identified on the basis of one or two-
component black body fits to far-IR and sub-millimeter data on galaxies spanning
a wide range of morphological types and levels of star-forming activity.
The literature can best be summarized by defining, in an ad hoc manner,
three dust components characterized by
their temperature. We define Very Cold Dust ( VCD ) as dust having $T_d 
< 15$ K; Cold Dust ( CD ) as $ 15 \leq T_d < 25$ K; and Warm Dust (WD) with
$T_d \geq 25$ K. The continuua of galaxies often appear to be dominated by one 
or more of these thermal components at far-IR and sub-millimeter wavelengths.

A preliminary study of our Milky Way using the COBE, Far Infrared Absolute
Spectrophotometer (FIRAS), (Wright et al. 1991) concluded that the spectral
data from 121 to 2600 \mic m imply a two-component continuum model
with a dust emission index $\nu^{\alpha}$, of $\alpha = 2.0$, and $T_{CD}$ = 20.4 and $T_{VCD} = $4.77 K. The 
VCD dust component, however,
contributed only 0.11 \% of the total dust emission. 
Sodroski et al. (1994) used  COBE, DIRBE all-sky data to 
show that a pervasive CD dust component exists with $17 < T_d < 
22$ K which can be detected at $60 < \lambda (\mu m) < 240 $. Indeed, this 
component is closely associated with the diffuse HI clouds and cold
molecular clouds that make up the far-IR `cirrus' discovered by IRAS
at 100 \mic m ( Soifer et al., 1984).

UKIRT observations of the nuclear regions of 11 IR-bright spiral galaxies by 
Eales, Wynn-Williams and Duncan (1989)
obtained fits to their 60 - 1100 \mic m continuua using a
single thermal component model with $\alpha = 2.0$. The dust temperatures 
providing the best fits were in the range from 30 - 50 K indicative of WD 
presumably associated with active star-forming regions.
A similar IRAS/sub-mm study by Clements, Andreani and Chase (1993) of 5
bright galaxies selected from the far-IR catalog by Smith et al. (1987) also
concluded that a single WD component yielded a superior fit to the
photometric data with $ 28 \leq T_d \leq 35$.
At least for luminous far-IR selected galaxies, the conclusion appears to be 
that a single WD
dust component dominates the far-IR emission at $\lambda \ge 60$ \mic m. 

Among the less active, late-type galaxies, a different population of dust 
grains appears to dominate the galactic emission.
Chini et al. (1986) investigated a sample of 18 
Sb-Sc spiral galaxies. Rather than a single WD component, they find
 two thermal components in the far-IR; one 
WD component typically has $T_d \approx$ 53 K, while a cooler 
CD component  has $T_d \approx$ 16 K. 
Eales, Wynn-Williams and Duncan (1989) included in their study several 
of the galaxies observed by Chini et al. (1986) but were unable to reach the
same conclusions, finding only conclusive evidence
for a single CD component.
The issue of one vs two thermal components for late-type galaxies
was re-opened by Chini and Kruegel (1993) in studies of NGC 660 and 
UGC 3490 who, again, identify two components : $T_{WD} \approx $ 50 K and 
$T_{CD} \approx$ 17 K in these galaxies.
They proposed that the sub-millimeter disagreements were the result of 
investigators not observing  galaxies 
with the same beam size (90$\prime\prime$), 
chopper throw, and wavelength as had Chini et al. (1986).

At the present time, the  controversy remains unresolved as to
 whether a sample of 
`normal' spiral galaxies distinguished only by their optical rather than far-IR 
luminosity have one or two thermal components dominating  their far-IR 
luminosity at $\lambda > 60$ \mic m. Thuan and Sauvage 
(1992) reviewed the results from a variety of
often conflicting surveys, and through a series of 
cross-correlations between the CfA Galaxy Catalog (Huchra et al., 1992; hereafter
 the `CfA Catalog')  
and the $H_{\alpha}$ survey of galaxies by Kennicut (1983), 
showed that the expected correlation between star-forming activity and total
far-IR 
luminosity does not occur in the case of `normal' spiral galaxies. Star-forming regions 
are apparently less important in determining the integrated
far-IR luminosity of a normal ( non-AGN, non-starburst) spiral galaxy 
than some other parameter of the galaxian interstellar medium. They suggest 
that CD grains associated with cirrus-like dust at $T_d \approx 20$ K 
may be the principle source of far-IR luminosity at $\lambda > 60$ \mic m.
 This
is consistent with the findings of Walterbos and Schwering (1987) and Rice et 
al. (1990) who showed that 85 \% of the far-IR emission from M31 (Sb-type) and 
40 \% of the far-IR emission from M33 (Scd) are associated with cirrus-type 
dust. Sodroski et al. (1994) show a similar trend for the Milky Way ( Sb-Sc)
with the Galactic plane cirrus component contributing as much as 90 \% of the far-IR
luminosity. 
 
Odenwald, Newmark and Smoot (1995; hereafter `Paper I')
reported the detection of 56 galaxies between 12 and 240 \mic m based
on a comparison of the IRAS Catalog of Extragalactic Objects (Fullmer and
Lonsdale, 1989; hereafter the 'IRAS Catalog) and the CfA Catalog, 
with the all-sky DIRBE survey. 
 DIRBE's large beam size of $0.7^o$ makes it an ideal instrument
for determining the integrated far-IR flux from an entire galaxy, not just its
bright nuclear regions. With the exception of M-31 and the Magellanic Clouds,
all galaxies bright enough to be detected by DIRBE will be observed as
unresolvable point sources so that all photometric measurements of
external galaxies are measurements of the integrated total emission. 
DIRBE's photometric overlap with the IRAS 12-100 \mic m
bands, and additional observations at 140 and 240 \mic m provide
important constraints to the strength of any dust components with temperatures
in the range from $15 < T_d(K) < 25$. Not all of the galaxies 
listed in Paper I were detected at 140
or 240 \mic m, but a significant subset of 7 galaxies were bright enough to
have measurable emission at either or both of these wavelengths.

In this paper, we continue an analysis of the DIRBE
galaxy sample by obtaining photometry for each galaxy in the DIRBE 12 - 240 
\mic m bands, and refining the original detection list in Paper I. 
We will then use the photometric data
to constrain as far as is feasible, the far-IR emitting WD, CD and VCD dust
components in the detected galaxies.

\vskip 0.2truein
\centerline{\bf 2.0 OBSERVATIONS}
\vskip 0.1truein

     Between December, 1989 and September, 1990, the 
DIRBE instrument on board NASA's
COBE satellite surveyed the entire sky in 10 photometric bands covering
the wavelength region from 1.25 to 240 \mic m.
A detailed description of the DIRBE instrument and the COBE mission is given
by Boggess et al. (1992). 
  An extensive discussion of the absolute calibration of the DIRBE photometry 
may be found in the COBE, `DIRBE Explanatory Supplement' ( COBE,1995), 
but for convenience we will briefly review
 some of the salient issues which affect galaxy photometry.

The 2-dimensional beam profiles in the 12-100 \mic m bands were determined by
measuring the intensity changes of numerous bright stars during the course of
the entire mission as they transited the beam. For the 140 and 240 \mic m 
bands, transits of the planets Jupiter and Saturn
 were used. The instantaneous
DIRBE beam profile in each band  is constrained by an internal field 
stop to $0.7^o \times 0.7^o$. 
The beam solid angles for the 12-240 \mic m bands are;
1.42, 1.48, 1.51, 1.44, 1.36 and 1.33 $\times 10^{-4}$ sr with uncertainties of
4, 5, 15, 12, 26 and 37 $\times 10^{-7}$ sr respectively. 

The COBE attitude control system provides information on the instantaneous
pointing direction of the satellite throughout the mission. This information
is provided by a combination of sun sensors, Earth horizon sensors, rate-
integrating gyros and magnetometers, and is used to establish
 an attitude solution for each detector as a function
of observing time. The rms uncertainty in the attitude solution is 0.93' based 
on sightings of specific calibration objects in the near-IR 1.25 - 4.9 \mic m
bands.

The DIRBE photometry has been corrected for detector gain variations due to  
environmental influences and instrument instabilities, and then calibrated to 
physical units.
Observations of approximately 144 celestial calibrators ( e.g. stars, planetary
nebulae, HII regions) are then used to establish a relative photometric
system which is stable over the duration of the mission. This relative
photometric scale is then converted to an absolute scale by observing a small
sample of well known discrete sources.

As described in the DIRBE Explanatory Supplement (\S 4.5.3.3)
the bright star Sirius was used to calibrate the 12 \mic m band; the planetary
nebula NGC 7027 was used for the 25 \mic m band; the 60 and 100 \mic m bands 
were calibrated by using the planet Uranus; and Jupiter was used for the 140 
and 240 \mic m bands. The quadrature sum of the contributions to the absolute
flux calibration in the 12-240 \mic m bands are (in percent): $C_F$ = 11.7, 
15.5, 8.6, 13.7, 10.1 and 10.1 respectively. The uncertainties in the absolute 
surface brightness calibration, $C_I$, are virtually identical. The primary
source of the uncertainty is in the adopted brightnesses of Sirius, NGC 7027 
and the planets Jupiter and Uranus. Also as explained in the DIRBE
Explanatory Supplement (\S 4.5.4.2 and Table 4.5-9)
a comparison of one band with the others for purposes of spectral fitting
and source color determination leads to uncertainties of $\approx$ 15 \%
depending on the band pairs considered.  

The initial detection search in Paper I was conducted using the cold mission,
averaged skymap data.
The positions searched for galaxy detections were
 obtained from the IRAS Catalog. We also
searched the CfA Catalog for detections in the 4\% of the sky not covered by
the IRAS survey which according to the CfA Catalog included
1251 optically-identified galaxies brighter than $+14.5^m$.
Initial estimates of the point source flux in the skymap data were obtained
by removing a simple 2-dimensional background model from the position of
each galaxy in the skymap.

In order to be considered a detection, in Paper 1 we described how the 
candidate had to be visible at a signal-to-noise (S/N) of at least 5.0 after
background removal  in a single band, or at a S/N exceeding
4 in at least two bands. In the latter case, the effective
S/N $> 4\sqrt{2} = 5.6$ matched or exceeded the single-band threshold.
The search was then considered to be complete to a uniform S/N = 5. 

As the IRAS survey discovered (see Low et al., 1989), much of the sky at
$\lambda > 60$ \mic m is covered by interstellar, infrared cirrus clouds which can
have considerable internal structure from a few arcminutes to many degrees
in scale. Large-beam observations of such a background can lead to
spurious detections of unresolved cirrus cloud features as well as
bona fide non-cirrus background sources.
If a candidate satisfied the S/N criterion, the surrounding $7^o \times 7^o$ field
was then inspected visually to 
insure that the candidate could not plausibly be assigned to an unrelated 
cirrus feature. 
We also consulted the ``Catalog of Infrared Observations'' (Gezari, Schmitz and 
Meade, 1993) to eliminate prominent, Galactic IR sources
which may coincide with the 3$\times$3 pixel DIRBE beam patch centered on 
the galaxy position. The initial search ( Paper I) for 
extragalactic detections produced 53 candidates in the IRAS Catalog, and 3 
candidates in the CfA Catalog. In what follows, we will subject this initial
detection list to a more rigorous analysis for spurious false detections,
and obtain a final list of DIRBE extragalactic detections. 

\vskip 0.2truein
\centerline{\bf 3.0 GALAXY DETECTIONS AND PHOTOMETRY}
\vskip 0.1truein
\centerline{\bf 3.1 The Final Catalog}
\vskip 0.1truein

In Figure 1, we show the results of the initial search described in Paper 1. 
Each image represents a $7^o \times 7^o$ field centered on the pixel coinciding 
with the cataloged galaxy position. The ordering of the galaxies is based
on the refined classification criteria we will discuss in this section.
A 2nd order background model has been 
removed from each field, and the pixel intensities have been
re-scaled, so that the intensity of the 
pixel nearest the galaxy position is normalized to 1.0. This particular image 
transformation, used only in the inspection for foreground clutter, optimizes
the contrast of the foreground clutter in the field. 
Table 1 gives the S/N of the peak galaxy emission found in each of the DIRBE
images shown in Figure 1, 
along with the total S/N for the candidate. For measurements near the
noise limit of the sky map, some of the listed S/Ns will be negative.
The total S/N is computed
from the square root of the  quadrature sum of the individual in-band S/Ns.

It is evident that many of the fields, especially those 
corresponding to the weakest detections such as NGC 1313 and NGC 3621, contain
significant structure at 60 and 100 \mic m. Indeed, many of the candidates
with  a total S/N $<$ 8  show enhanced foreground clutter.
The most important question we must now address is, given the large size
of the DIRBE beam and the complexity of the cirrus emission, how may we
insure that a particular candidate is a galaxy rather than a detection of a
bright cirrus feature? 

In Paper I, we utilized
a simple S/N detection criterion to establish the candidate list, followed
by a visual
inspection of each candidate field to eliminate objects that were clearly in
fields with a complex background.
For example, the  cometary 
star-forming region NGC 5367 which is a bright, compact far-IR source 
according to Odenwald (1988) appeared to coincide with the CfA galaxy
A1356-3949. This would have resulted in a false detection of this galaxy in the
DIRBE skymap data. We would now like to investigate improving
the initial survey described in Paper I 
by utilizing a group of indices designed to characterize
the presence of cirrus emission in each field.

We present in Table 2 the revised catalog of 54 DIRBE galaxy candidates
based on the results from Paper I. We do not include in this study either
the Large or Small Magellanic Clouds since they were resolved by DIRBE.
Excluded from this table is the nearby galaxy M-31 which DIRBE also
detected as an extended infrared source. This new tabulation, moreover,
does not list  the previously identified DIRBE candidate `NGC 7625'
which was found to be the galactic source CRL 3068 located 18$^{\prime}$ 
from the galaxy position. 
Column 1 is the galaxy name; column 2 indicates the morphological type;
columns 3-4 give the Galactic latitude and longitude of the galaxy
based on its IRAS or CfA Catalog positions.
The degree of cirrus contamination for each galaxy field is indicated by
a set of quality indices appearing in columns 5 - 12. 
The first four indices were obtained from the IRAS Point Source Catalog and
are generally
recognized as useful cirrus flags for studies of point source in the IRAS
data. SES1 counts, in each band, the number of hours-confirmed small extended 
sources within 6$^{\prime}$ of the identified point source. SES2 represents a similar
tally for weeks-confirmed sources. According to the IRAS Explanatory 
Supplement ($\S$ VII-36.H.1.b), both indices warn of the presence of 
extended, Galactic foreground structure. Columns 5 and 6 
present the SES1 and SES2 'small extended source' confusion flags 
at 100 \mic m. Values less than 2 are considered to be indicative of
point sources
in `clean' fields. In columns 7 and 8 we present the IRAS CIRR1 and CIRR2 
indices. CIRR1 gives the total number of sources detected at 100 \mic m
in a 1 square degree box centered on the IRAS point source. Values exceeding
3 indicate significant cirrus or extended emission. CIRR2
is the logarithmic ratio of the point source 100 \mic m flux to the cirrus
flux from spatially-filtered 0.5$^o$ IRAS data. 
Values exceeding 3 also indicate significant cirrus contamination.
The CIRR1 and 2 indices are well-matched to the DIRBE beam and will be given
higher weight than the comparatively small angular scale ( $\approx 6'$) SES1 
and SES2 indices.

One of the difficulties in using the IRAS cirrus flags as a final
arbiter of whether or not the candidate is a cirrus feature  is that
objects such as M33, M82 and even NGC253 have large CIRR1 indices.
In a fully automated search for detections, these galaxies would be overlooked
in favor of other fainter 
candidates such as NGC 4102 or NGC 4038 which have lower
cirrus indices. This suggests that
the IRAS-based SES and CIRR indices cannot constitute the entire basis for 
deciding between a detection of clutter and the detection of a galaxy in the
DIRBE skymap data.
We have, therefore, defined three new indices appropriate for DIRBE skymap
data which provide a measure of the underlying structure of the foreground
against which each galaxy is seen.
 N3x3, presented in column 9 of Table 2,
represents the number of IRAS point sources found in the $3 \times 3$ pixel
beam patch centered on the galaxy.
 This index is similar to the IRAS CIRR1 index, but
includes all IRAS point sources, not just those detected only at 100 \mic m. 
For IRAS-detected galaxies its minimum value is unity. Values larger than this can
occur if the galaxy is resolved by IRAS, or if the
IRAS field contains a number of bright, discrete features in the infrared
foreground.
We define a second index, N100, as the number of 100 \mic m IRAS sources
(excluding the candidate galaxy) that were
counted in N3x3, and  which are brighter than 10 \% of the galaxy's 
peak IRAS flux appearing in the IRAS Point Source Catalog. 
The final index, RB, is the ratio of the rms background 100 \mic m 
intensity of the DIRBE skymap region within $2^o$ of the
galaxy (but not including the 3x3-pixel galaxy patch),  and
the rms  of a sky patch within $2^o$ of the North Galactic
Pole. After the removal of a smooth background gradient, 
the NGP region was found to have an rms intensity of 0.72 MJy/sr at
100 \mic m. Values for RB near unity 
indicate that the galaxy is situated in a region with essentially the same
degree of detectable clutter at a scale of a few degrees as compared with the
reference region near the North Galactic Pole. Values near RB = 0 indicate that
the galaxy region has less clutter than even the North Galactic Pole.
In column 12, we also present the total S/N for the galaxy candidate
obtained from a quadrature sum of the individual S/Ns for each of the 6
bands we surveyed.

Comparing the various confusion flags in Table 2 against the fields in Figure
1, we see no simple correlation between detectability and
background clutter without exempting a number of galaxies from the
analysis. The CIRR1 and
CIRR2 indices yield more clutter discrimination at the scale of the DIRBE beam
and show that in addition to M82 and M101, the galaxies M33, NGC 1313,
NGC 3256 and NGC 4945 have values exceeding 3. These are, therefore, suspect
of having cirrus contamination and would have to be eliminated as detections
in a conservative cataloging of DIRBE galaxy candidates. 

The collection of indices, N3x3, N100, RB, SNR, derived from the DIRBE
and IRAS data yield a better measure of the amount of clutter in each field,
in particular, the RB and total S/N indices. 
From a comparison of Figure 1 with the RB index in Table 2, column 11,
 and the total S/N for the candidate in column 12, 
we propose the following algorithm for reducing false detections due
to cirrus or other foreground clutter in our sample.

All candidates with S/N $>$ 15 will be considered positive detections. This
yields a sample of 17 galaxies which we identify as Group A. With the 
exception of M82, NGC 4945, NGC 5128 and IC 342, the Group A
galaxies have RB $\leq$ 3 and CIRR2 $\leq$ 3,
both indicating minimal background cirrus clutter in the immediate vicinity
of each galaxy. The large RB value for M82 is the
result of the presence of a bright cirrus
feature near M82 which is apparent in Figure 1. 
The galaxies  NGC 1808 and IC 342, on the other hand,
 have the most cluttered fields of the galaxies in this group. Their 
index sets (SES1, SES2, CIRR1, CIRR2, RB)  are  (0,0,1,1,2) and
(4,1,0,3,15) respectively, and indicate that only IC 342 has a
significant probability of being a false detection of cirrus.

We define Group B candidates as detections which have $8 < S/N < 15$, and
as for Group A, 
RB $\leq$ 3. This sample of 17 galaxies has SES2 and CIRR2 $\leq$ 2
indicating relatively
low cirrus contamination. At the scale of the DIRBE beam, and over
a region of 7$^o$, Figure 1 shows that unlike Group A, and despite their low
cirrus indices in Table 2, some of the fields, nevertheless,
contain some clutter, particularly NGC 4490 and NGC 613.

The third, Group C,  represent the
fainter candidates in this survey and have $ 5 < S/N < 8$. Of these,
NGC 891, NGC 1097A, NGC 3256, NGC 3621, NGC 5128 and IC 751,
are adjacent to
bright regions of cirrus emission and unlike M82, have insufficient
S/N to readily distinguish them from bright cores in the cirrus emission other
than by virtue of their coincidence with the galaxy positions. NGC 891,
NGC 1097A and NGC 5128, nevertheless, represent the more plausible galaxy 
detections among these 6 candidates.

From the previous analysis, we conclude that the 39 members
of Group A and B constitute the most reliable extragalactic detections
obtained from the DIRBE skymap data at 12-240 \mic m.

\vskip 0.2truein
\centerline{\bf 3.2 Point Source Photometry}
\vskip 0.1truein

 The DIRBE skymaps have been 
optimized for preserving the photometric accuracy of extended emission at 
scales larger than the DIRBE beam. The time-ordered data is the preferred
data to use to obtain accurate point source photometry, however, this is a 
time consuming procedure beyond the scope of the present study.
Point source photometry used in this study is derived from the cold,
 mission-averaged
DIRBE skymap data. A correction factor is applied to the peak emission detected
towards a galaxy to reflect the true beam response.

Due to the pixelization procedure, the peak IR emission from a galaxy or other
unresolved source identified in the skymap data
may be up to $\pm 1/2$ pixel from its nominal optical position.
Moreover, since the DIRBE beam response function is not flat over its full field of
view, this can cause the galaxy emission peak detected by DIRBE to be as much as
1 pixel different from the true galaxy position. We have compensated for these
effects by calculating the response of the DIRBE beam in each band at the location
of the galaxy rather than at the skymap position by using an azimuthally -
averaged template of the beam response computed from many transits of calibrator
sources across the beam ( Odegard, 1995). 
From the beam template at the position of the optical
galaxy, we determined the responsivity of the detector, $R_b$, and then
used the skymap flux density at this pixel, $S_p$, to obtain a corrected peak
flux density, $S_c = S_p/R_b$. Typically, $0.9 < R_b < 1.0$ so the correction
was, generally, less than 10\%.

The results of this photometric analysis are presented in Table 3.
Column 1 is the galaxy name; column 2 is the galaxy type; column 3
is the maximum B-band  angular size in arcminutes of the
optical galaxy based on the tabulations of $D_{25}$ by de Vaucouleurs et al. (
1976, 1991); columns 4-9 give the background-subtracted and beam response
corrected DIRBE 
peak flux densities, $S_c$, in Janskys or their 3-$\sigma$ upper limits.
A color correction must be applied to the photometric values in
Table 3 if the source  effective temperatures
are different from that of the DIRBE absolute calibrators. Tables of these
correction factors may be found in the DIRBE Explanatory Supplement (
e.g. Appendix B).
As we can see in column 3, the maximum angular sizes of the galaxies are, with
few exceptions, smaller than the 42$'$ DIRBE beam. For the purposes of galaxy 
photometry, this large beam size implies that all but the nearest galaxies 
( e.g. M-31, the LMC and the SMC) will remain unresolved. 

\vskip 0.2truein
\centerline{\bf 4.0 CONTINUUM FITTING}
\vskip 0.1truein

We assume that the dust emissivity 
 follows the canonical $\nu^{\alpha}$ law with 
$\alpha$ = 2.0 ( see Eales et al., 1989; Wright et al., 1991). For a
galaxy for which photometry exists in two or more bands,
the photometry can be fit to obtain estimators for the 
dust temperature, $T_d$, and the optical depth, $\tau_{100}$, at 100 \mic m 
by using a dust emission function of the form:

$$I(\nu) = \tau_{100} ({\nu}/{\nu_{100}})^{2.0}  B( \nu, T_d ) $$

where  B($\nu, T_d$) is the
Planck function evaluated at a temperature $T_d$ and a frequency $\nu$.
 Color corrections have been  applied to the DIRBE photometric
data since according to the COBE Explanatory Supplement (1995), 
these corrections (C), defined as $I_{true} = {I_{observed} /  C}$,
  are significantly different from unity between 12  - 240 \mic m. 
The color corrections were determined as follows:
The spectral fits were first determined without the corrections.
A mean color correction was
then applied  representative of the average fitted temperature.
For  a dust grain index of  $\alpha$ = 2.0, the mean temperature of the entire
galaxy sample was found to be  $ <T_d> = 30.2$ K. The 
dust model fits were then recomputed for the color-corrected DIRBE fluxes to
obtain the final fitted dust temperatures and optical depths for each galaxy. 
The use of color-corrected DIRBE fluxes leads to no significant change in the 
resulting fitted dust temperatures, but does produce  an 
$\approx$ 20 \% increase in $\tau_{100}$ compared 
to the uncorrected dust model fits, which is significant when
estimating galactic dust masses.
The result of this single-component dust fitting is presented in Table 4.
Columns 1 and 2 are the galaxy name and morphological class;
 columns 3 and 4  give the fitted temperature and its 
uncertainty.

The galaxies in Table 4 have been segregated according to their 
morphological types so that any trends in the dust temperatures could be more 
easily recognized. The preponderance of the galaxies are late-type spirals,
and systems identified as `active' or `peculiar' either as Seyferts, LINERS,
or tidally - interacting.
Based on the fitted temperatures and their uncertainties,
we can derive a weighted, mean temperature for each class
defined as, $ < T >_w = \Sigma (T_i / \sigma_i^2 ) / \Sigma (1/\sigma_i^2) $
with $< \sigma >_w^2 = 1/\Sigma(1/\sigma_i^2) $ ( see Bevington, 1969).
The weights, $\sigma_i$, were determined from the individual black body fits
to each galaxy's photometric data.

 The mean, weighted dust temperatures 
span a range from 27 - 35 K, with specific class temperatures of $27.6 \pm 0.1$
K for the spirals, $26.7 \pm 0.3$ K for the barred spirals, and $35.5 \pm 0.1$
K for the active galaxies. The uncertainties are biased in favor of the
small number of galaxies in each cohort which have the lowest temperature
 uncertainties. The spirals and barred spirals do not differ significantly
from one another, however, the active galaxies are significantly warmer.  

The galaxies form two groups in which
a WD component is dominant for the AGN and starburst-type systems, and where
a CD component appears to be dominant in the remaining non-AGN/starburst galaxies.
This is, of itself, not a new result, but is corroborative of
findings by other investigators when similar categories of galaxies are
compared. 

\vskip 0.2truein
\centerline{\bf 4.1 Dust Masses}
\vskip 0.1truein

Using the single thermal component fits as an initial model, we can investigate
whether significant VCD may also be present in these galaxies based on the
upper limits or detections at 140 and 240 \mic m. 
Black body dust grains with $T_d = 15 K$ have their 
peak emission at 245 \mic m and contribute only 25 \% of their peak emission
at 100 \mic m, which means that the DIRBE 240 \mic m band is optimally placed to measure
the strength of this thermal component.
Assigning a temperature of
15 K to this VCD component, and subtracting the fitted, one-component
 thermal model from
the measured continuum, we can estimate from the residual 240 \mic m emission
or its upper limit, a mass limit for this VCD
that would be consistent with the galaxy continuua as measured by DIRBE.

For dust grains with an emissivity coefficient, 
$Q_\nu \propto \nu^{1}$, the dust mass based on the
galaxy's measured infrared flux density, the estimated dust temperature, and its
distance is
given by  Knapp (1991) and Clements, Andreani and Chase (1993) as,
$$M_d = 4.8 S_{100} D^2 (exp(144/T) - 1)$$where $S_{100}$ is the residual flux density, or its upper limit, 
at 100 \mic m in units of Janskys, D is
the distance to the galaxy in megaparsecs, and T is the assumed dust 
temperature. This can be rescaled for dust grains with $Q_\nu \propto
\nu^{2}$ as,

$$M100 = 3.4 S_{100} D^2 (exp(144/T) - 1)$$

For purposes of estimating the mass of the emitting cirrus-like dust 
in each galaxy,
which we define as M100, we use the temperature for the far-IR emission 
obtained from our single-component spectral fits. Table 4 presents the
derived dust properties obtained in this way for the galaxies in Table 3 with
known distances. In column 3 and 4 the fitted
temperature and its uncertainty are given; column 5 is the adopted distance
to each galaxy obtained from the listings by Huchtmeier and Richter (1989),
and the redshifts tabulated by Spinoglio and Malkan ( 1989) with $H_0$ = 75
km sec$^{-1}$ Mpc$^{-1}$
; and column 6 is the estimated mass of the fitted dust component.

We also provide upper limits to the mass of a potential VCD component in each
galaxy with an adopted temperature of 15K by computing the difference between
the DIRBE photometry at 240 \mic m, and the predicted level of the fitted
thermal component detected at shorter wavelengths. The relevant relation between
the far-IR emission limit at 240 \mic m and the effective dust mass, M240, is given
by,

$$M240 = 270 S_{240} D^2 (exp(59/15) - 1)$$

The resulting upper limits are presented in Table 4, column 7, together with
the distance-independent ratio of M240/M100 in Column 8.

We see that the DIRBE observations provide  modest  constraints
on the magnitude of a hypothetical 
15 K VCD component in M33, M63, M83, NGC 253 and NGC 4945 on the basis
of direct detection of the far-IR continuum level for these galaxies at
140 and/or 240 \mic m. This component may not be larger in total
emitting mass than about 5 - 7 times the
detected CD component at 100 \mic m for M63 and M33, and not more than about 30 times
the detected CD components in  NGC 253, NGC 4945 and M 83. 
These galaxies appear to be well-fit by a single thermal component 
at $\lambda > 100$ \mic m. This strongly suggests
 that there are not large quantities of
dust within these galaxies that are coupled together by a large-scale thermal
gradient between 30 K - 15 K. This is analogous to what is observed in the 
Milky Way in which the dominant component is the 17 - 22 K cirrus dust
emission, with no other significant reservoir of dust detectable by the 
COBE FIRAS experiment in the range from 17 - 10 K. We note, however, that VCD 
near 5 K could be present with significantly more mass in the detected galaxies than
indicated by our upper limits for the hypothetical 15 K component, and still
remain undetectable by DIRBE at 240 \mic m.

For the remaining
galaxies not detected at 240 \mic m, the limits are considerably less
restrictive and in many cases can accomodate as much as 100 times as much dust
by mass at temperatures near 15 K as at 20 - 30 K. Sub-millimeter observations
of these galaxies are clearly required to improve these restrictions to
astrophysically interesting levels.

\vskip 0.2truein
\centerline{\bf 5.0  SUMMARY}
\vskip 0.1truein

From an initial catalog of 56 candidates, we present refined catalogs
of galaxies detected by DIRBE. The sample is dominated by spiral-type
galaxies with effective dust temperatures similar to Galactic cirrus 
emission. This shows that dust similar to Galactic cirrus is at least
thermally a common ingredient of late-type galaxies such
as the Milky Way.

An investigation of the continua of these galaxies show them to be
consistent with a single WD-like dust component which has its peak
emission near 100 \mic m, with temperatures from 25 - 35 K. Galaxies identified
as active or morphologically peculiar are systematically warmer than the
late-type spirals which dominate this sample.
A comparison of the mass ratios of a possible 15K dust component
to the warmer 20-25 K dust in the DIRBE-detected galaxies 
indicates that 15 K dust can exist in many of these galaxies in
quite large amounts compared with the detectable cirrus-like components seen
in our own Milky Way by DIRBE and IRAS.

\vskip 0.2truein
\centerline{\bf ACKNOWLEDGMENTS}
\vskip 0.1truein

The authors gratefully acknowledge the efforts of the DIRBE data processing
and validation teams in producing the high-quality datasets used in this 
investigation.  COBE is supported by NASA's Astrophysics Division. Goddard
Space Flight Center, under the scientific guidance of the COBE Science
Working Group, is responsible for the development and operation of COBE.

\newpage
\centerline{\bf Figure Captions}

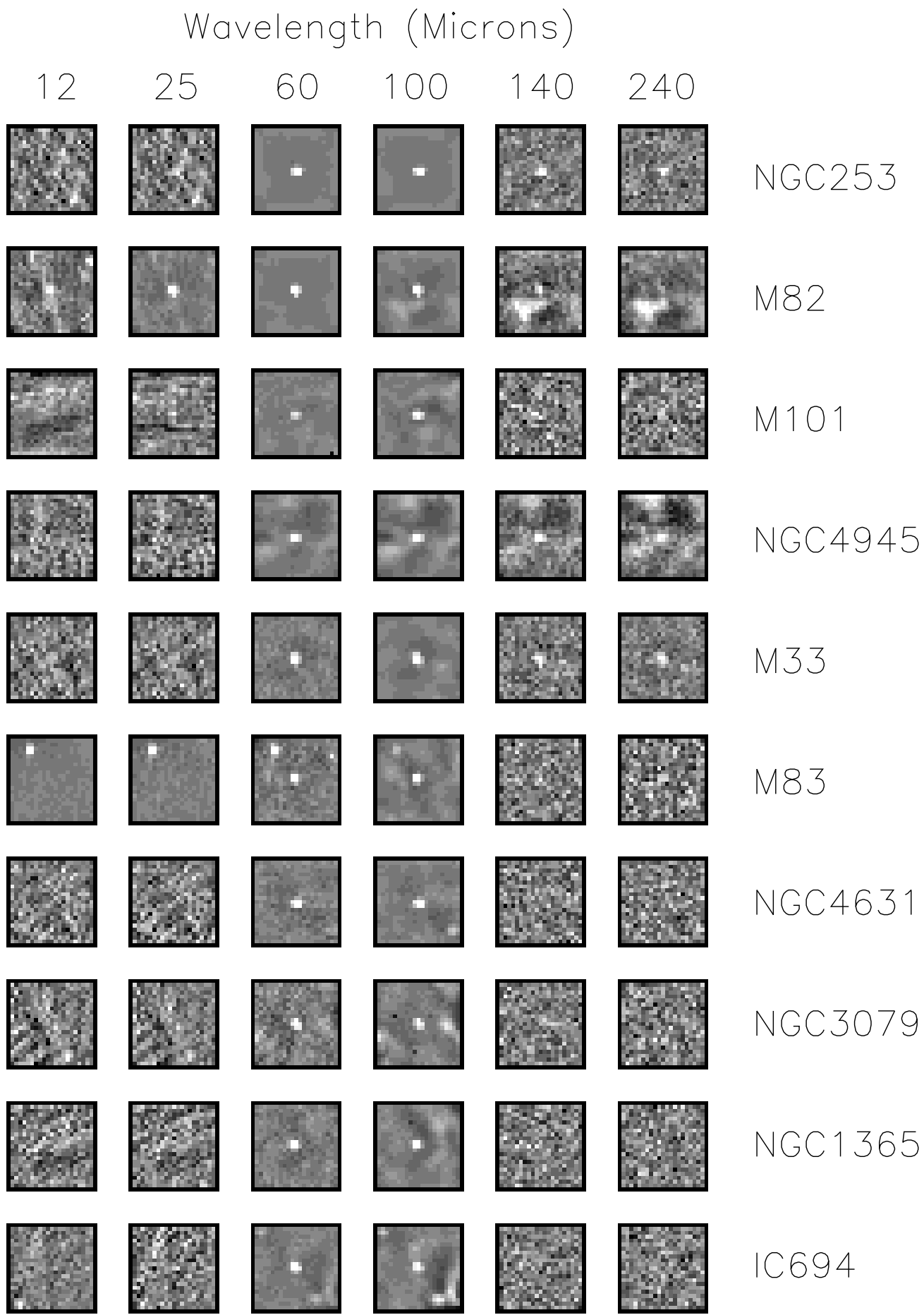
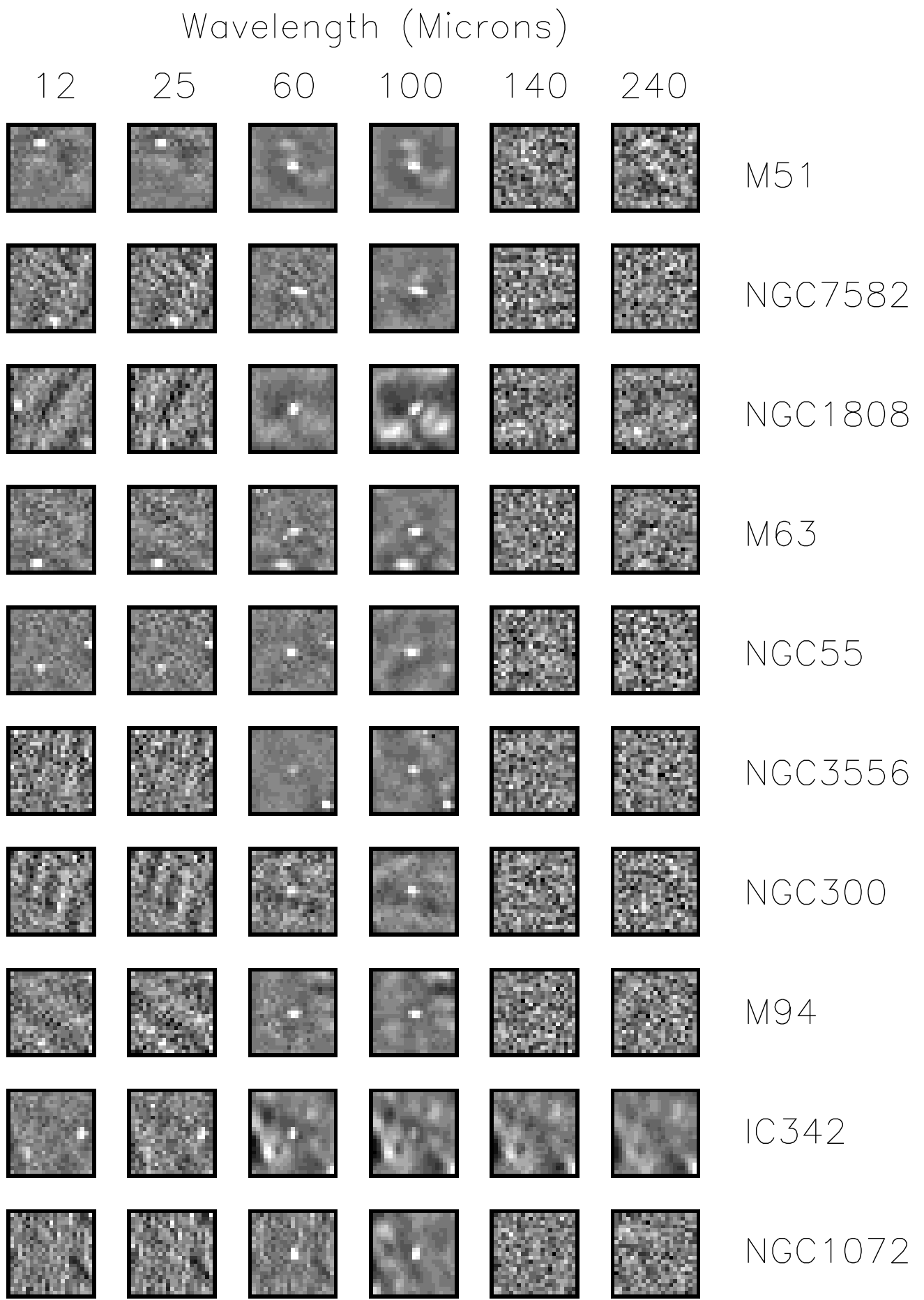
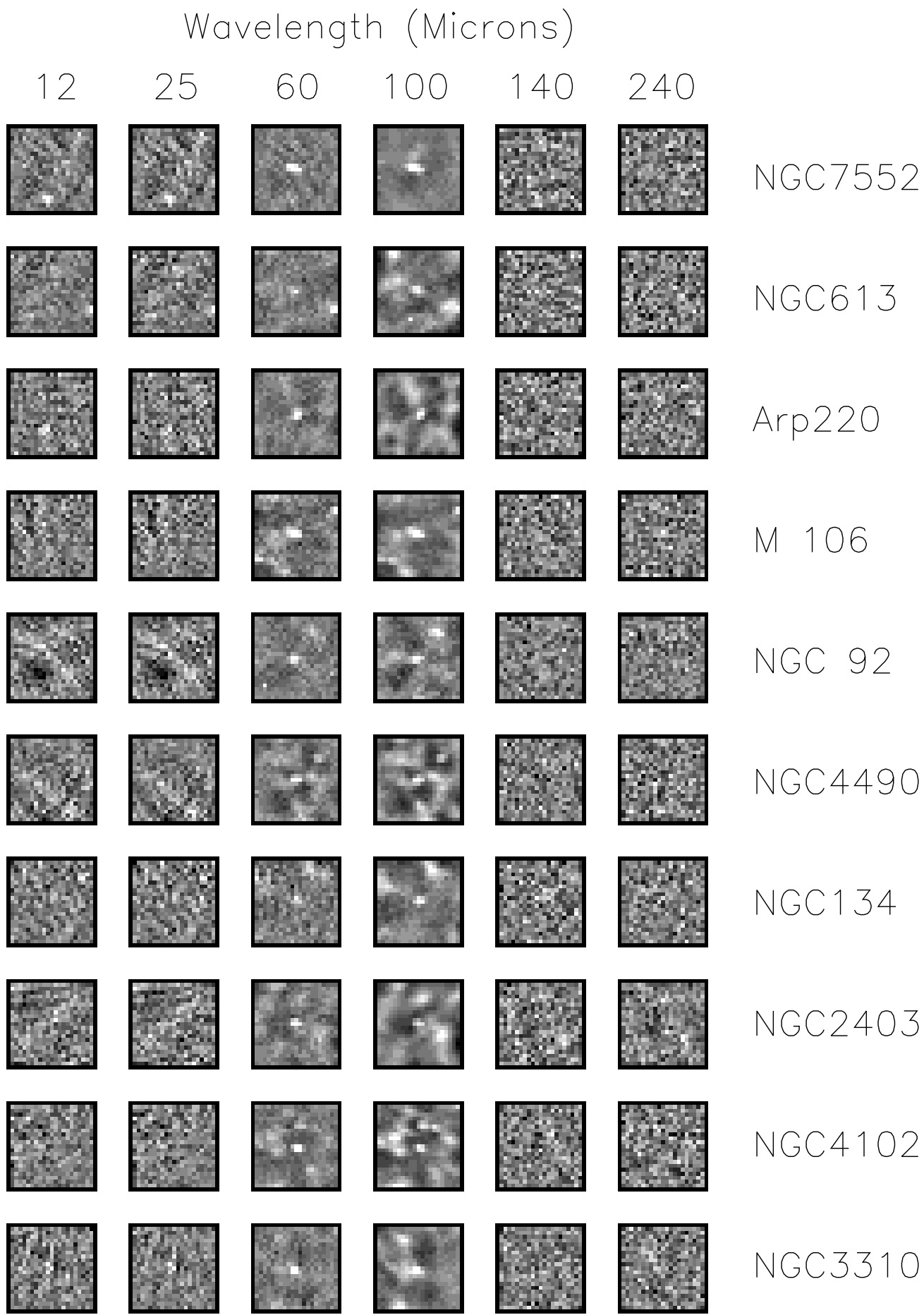
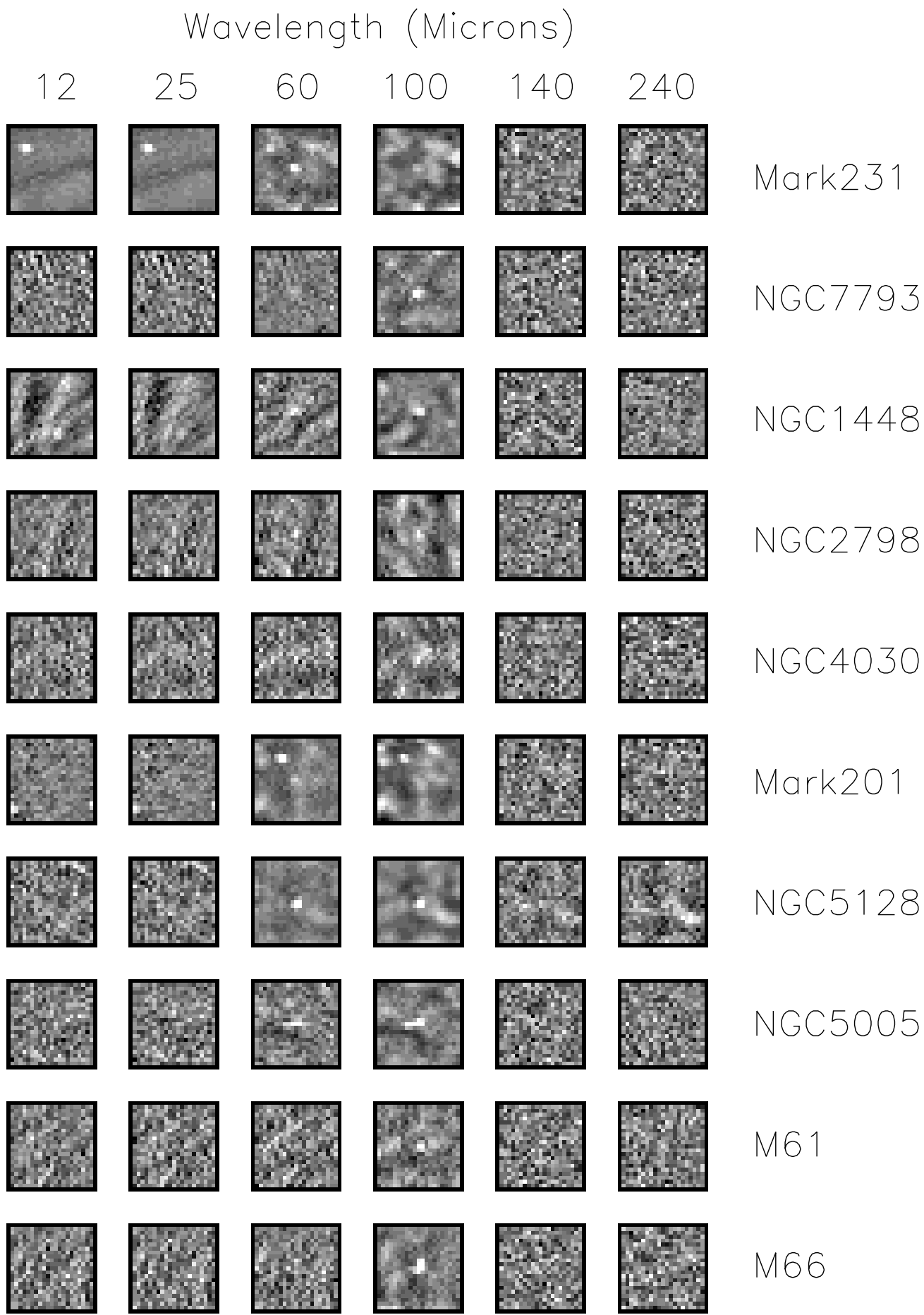
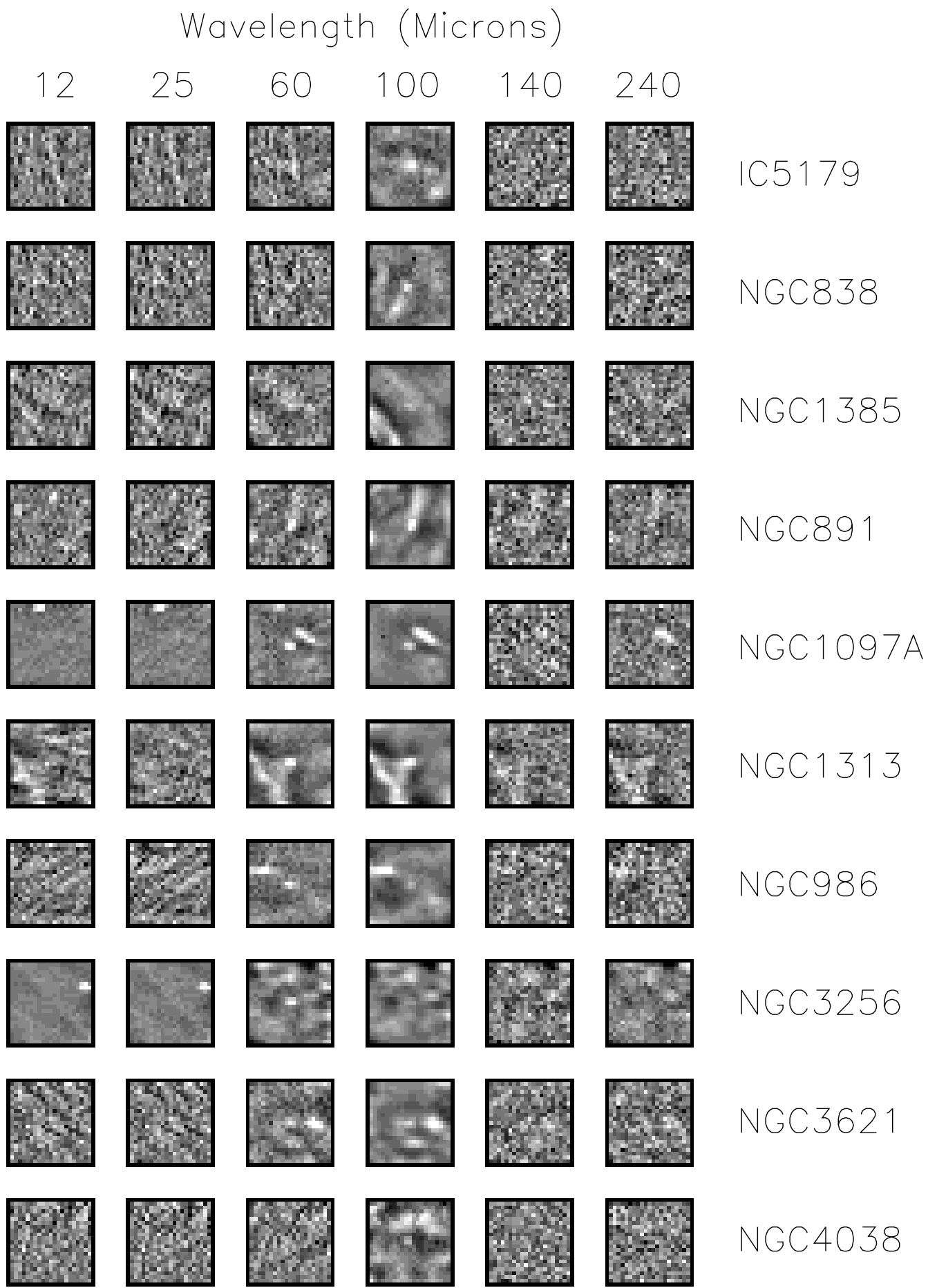
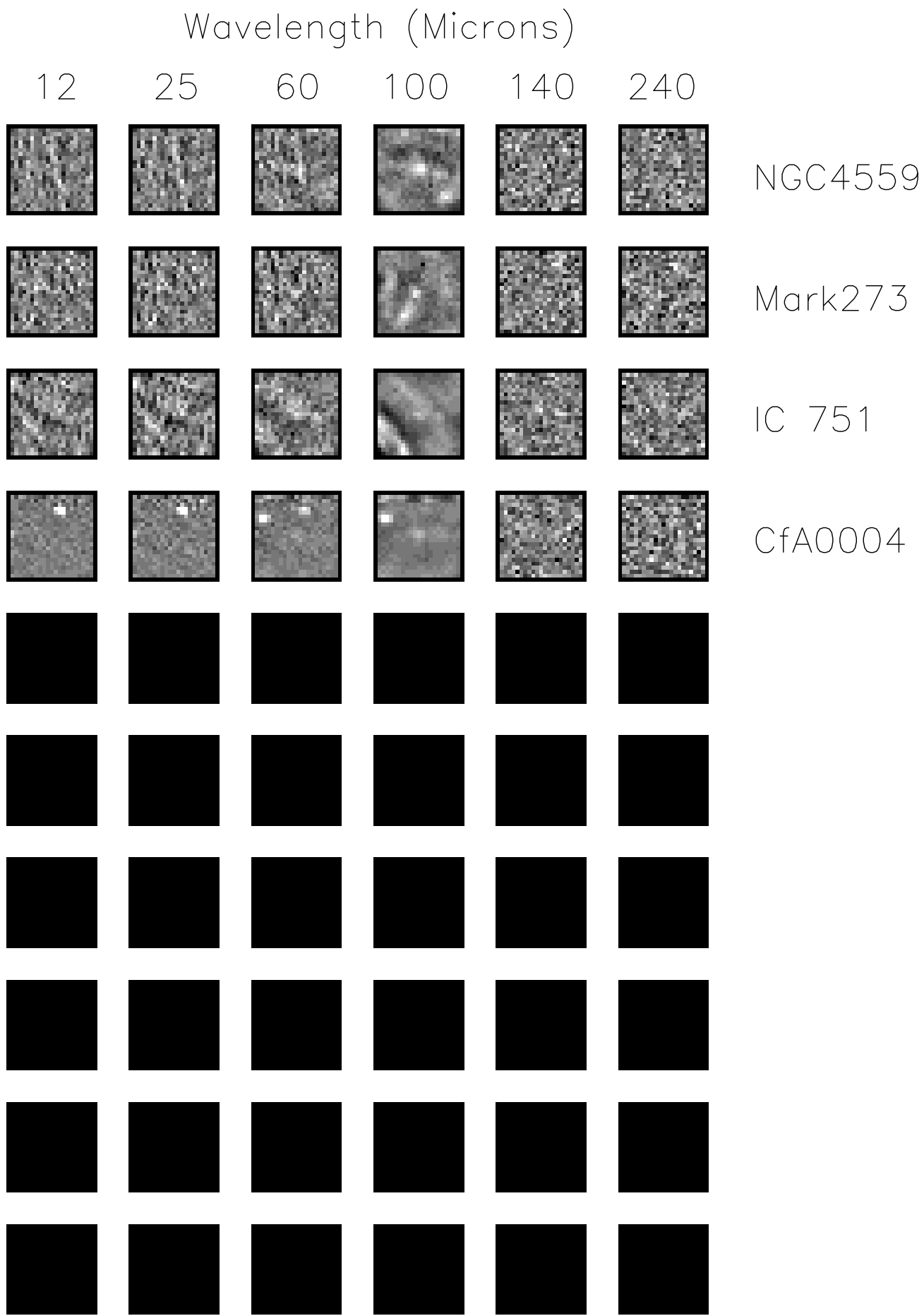
{\bf Figure 1}
\vskip 0.2truein

This figure presents the sky regions surrounding each of the DIRBE
galaxies identified in Table 1. The field of view in each band is 21$\times$21
pixels at a resolution of 21$\prime$/pixel. The images in each band have been
logarithmically stretched to facilitate the characterization of the
degree of background clutter. The galaxies are grouped following their entries
in Table 1 in which the S/Ns for the galaxy in each band is also presented. 

\vskip 1.0truein
{\bf Figure 2}
\vskip 0.2truein

The fitted spectra of the galaxies described in Table 4. Upper limits for
the DIRBE photometry are defined as 3 times the residual background emission
following the removal of a planar background model. The photometric
uncertainties for the measured flux densities are the 1-sigma total
photometric error defined as the rms quadrature sum of the absolute
photometric uncertainty and the background fitting uncertainty. The dashed and 
solid curves show the continuua produced from dust models involving emissivity
laws of $\nu^{2.0}$ and $\nu^{1.5}$ respectively.

\newpage
\centerline{\bf References}
\def\pp{\parshape 2 0truecm 16truecm 0.7truecm 15truecm}
\parindent 0pt
\pp

\pp    Bevington, P. R. 1969, `` Data Reduction and Error Analysis for
the Physical Sciences'', (McGraw-Hill Book Co.), p.130.

\pp    Boggess, N. et al. 1982, ApJ 397, 420.

\pp    Chini, R. et al. 1986, AA 166, L8.

\pp    Chini, R. \& Krugel, E. 1993, AA 279, 385.

\pp    Clements, D.L., Andreani, P. \& Chase, S.T. 1993, MNRAS 261, 299.

\pp   COBE Diffuse Infrared Background Experiment (DIRBE) Explanatory 
Supplement, eds. M.G. Hauser, T. Kelsall, D. Leisawitz, and J. Weiland,
COBE Ref. Pub. No. 95-A (Greenbelt, MD:NASA/GSFC).

\pp   Cox, P. \& Mezger, P.G. 1989, Astro. \& Astrop. Rev. 1, 49.

\pp    Desert, F.-X., Boulanger, F. \& Puget, J.L. 1990, AA 237, 215.

\pp    Eales, S.A., Wynn-Williams, C.G. \& Duncan, W.D. 1989, ApJ 339, 859. 

\pp    Fullmer, L. \& Londsdale, C. 1989, NSSDC Archive Number 7113.

\pp    Gezari, D.Y., Schmitz, M., Pitts, P.S. \& Mead, J.M. 1993,
``Catalog of Infrared  Observations'', NASA Reference Publication 1294

\pp    Hildebrand, R. et al. 1985, Icarus, 64, 64.

\pp    Huchra, J.P. et al. 1992, NSSDC Archive Number 7144.

\pp    Huchtmeier, W.K. \& Richter, O.G. 1989, ``A General Catalog of
HI Observations of Galaxies'', (Springer-Verlag, NY).

\pp    IPAC Users Guide, Edition 4, 1989, prepared by L. Fullmer et al., 
(JPL, Pasadena).

\pp    Kennicutt, R.C. 1983, ApJ 272, 54.

\pp    Knapp, G.R. \& Patten, B.M. 1991, AJ 101, 1609.

\pp    Low, F. et al, 1989, ApJ (Letter) 278, L19.

\pp    Mathis, J. Mezger, P. \& Panagia, N. 1983, AA 128, 212.

\pp    Odegard, N. 1995, Private Communication.

\pp    Odenwald, S., 1988, ApJ 325, 320.

\pp    Odenwald, S., Newmark, Jeffery, \& Smoot, George, 1995, 
IAU Symposium 168; April 24-26, 1995, University of Maryland.

\pp    Rice, W. et al. 1990, ApJ 358, 418. 

\pp    Smith, B.J. et al. 1987, ApJ 318, 161.

\pp    Sodroski, T. et al. 1994, ApJ 428, 638.

\pp    Soifer, B.T. et al. 1989, AJ 98, 766.

\pp    Spinoglio, L. \& Malkan, M.A. 1989, ApJ 342, 83.

\pp    Thuan, T.X. \& Sauvage, M. 1992, in ``Physics of Nearby Galaxies:
Nature of Nurture?'', eds. T.X. Thuan, C. Balkowski and J.T.T. Van, (Editions
Frontieres: Gif-sur-Yvette Cedex, France), p. 111.

\pp    Walterbos, R.A.M. \& Schwering, P.B.W. 1987, AA 180, 27.

\pp    Wright, N.L. et al. 1991, ApJ 381, 200.

\end{document}